\documentclass[aps,preprint,tightenlines,a4paper,showkeys,showpacs]{revtex4}%
\usepackage{amsfonts}
\usepackage{amsmath}
\usepackage{amssymb}
\usepackage{graphicx}%
\providecommand{\U}[1]{\protect\rule{.1in}{.1in}}

\begin{document}
\title{Microstructured superhydrorepellent surfaces: Effect of drop pressure on
fakir-state stability and apparent contact angles.}
\author{L. Afferrante and G. Carbone}
\affiliation{DIMeG Politecnico di Bari, v.le Japigia 182, 70126 Bari, Italy}
\affiliation{CEMeC, Politecnico di Bari, via Re David 200, 70125 Bari, Italy}

\begin{abstract}
In this paper we present a generalized Cassi-Baxter equation to take into
account the effect of drop pressure on the apparent contact angle
$\theta_{app}$. Also we determine the limiting pressure $p_{W}$ which causes
the impalement transition to the Wenzel state and the pull-off pressure
$p_{out}$ at which the drop detaches from the substrate. The calculations have
been carried out for axial-symmetric pillars of three different shapes:
conical, hemispherical topped and flat topped cylindrical pillars.
Calculations show that, assuming the same pillar spacing, conical pillars may
be more incline to undergo an impalement transition to the Wenzel state, but,
on the other hand, they are characterized by a vanishing pull-off pressure
which causes the drop not to adhere to the substrate and therefore to detach
very easily. We infer that this property should strongly reduce the contact
angle hysteresis as experimentally osberved in Ref.
\cite{Martines-Conical-Shape}. It is possible to combine large resistance to
impalement transition (i.e. large value of $p_{W}$) and small (or even
vanishing) detaching pressure $p_{out}$ by employing cylindrical pillars with
conical tips. We also show that depending on the particular pillar geometry,
the effect of drop pressure on the apparent contact angle $\theta_{app}$ may
be more or less significant. In particular we show that in case of conical
pillars increasing the drop pressure causes a significant decrease of
$\theta_{app}$ in agreement with some experimental investigations
\cite{LafunaTransitio}, whereas $\theta_{app}$ slightly increases for
hemispherical or flat topped cylindrical pillars.

\end{abstract}
\maketitle

\section{Introduction}

\begin{figure}[ptb]
\begin{center}
\includegraphics[
height=7.38cm,
width=12.99cm
]{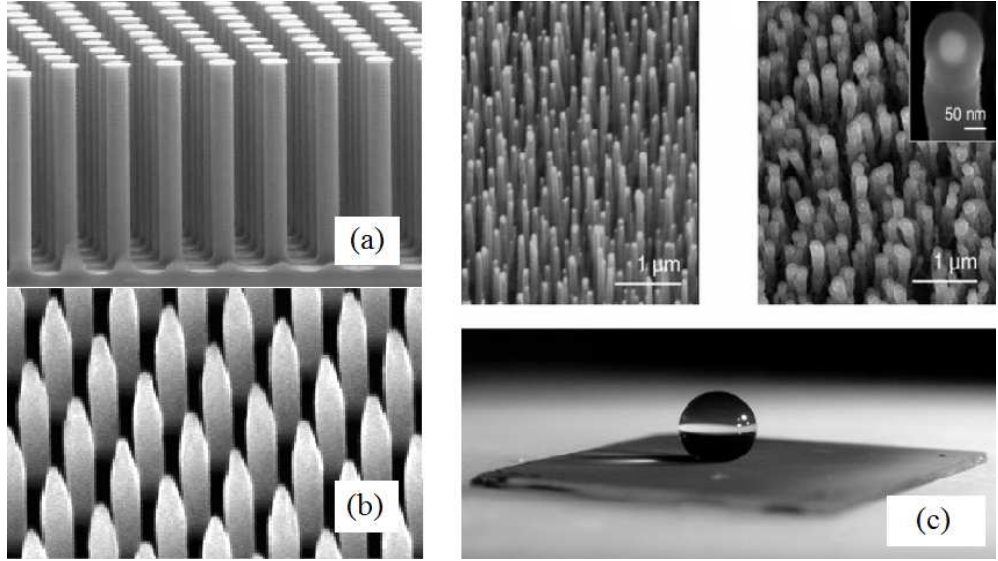}
\end{center}
\caption{A few examples of artifical superhydrorepellent surfaces. (a) A
uniform array of very slender cylindrical pillars with a flat tip (adapted
from Ref. \cite{callies-cylinders}), (b) a uniform array of cylinders with
tapered shaped tips (adapted form Ref. \cite{Martines-Conical-Shape}), and (c)
an example of superhydrorepellent carbon-nanotube forest (adapted from Ref.
\cite{nanotube-forest})}%
\label{Fig_SuperficiArtificiali}%
\end{figure}

Roughness-induced hydrophobicity is a well known effect observed in many plant
leaves, e.g. Sacred Lotus leaves (\textit{Nelumbo nucifera}) \cite{Barthlott},
and in other biological system as water striders \cite{water striders}
(\textit{Gerris remigis}) or mosquito (\textit{Culex pipiens}) eyes
\cite{Gao-Mosquito}. In all such cases the surface asperities make the liquid
able to be suspended on the asperity tips, resulting in a very large contact
angle (CA). The superlative water-repellency of such natural surfaces would be
very appreciated in many micro- and macro- engineering applications, as liquid
drops on super-hydrophobic surfaces may be very easily moved from one position
to the other by simply applying an external force, resulting in the
possibility to create a chemical microreactor and microfluidic microchips
\cite{Hermingauschip}, \cite{Washizu}. Beside these high-tech applications,
there is a strong interest of engineers in developing new commercial products
as self-cleaning paints and glass windows \cite{Blossey}, or super-hydrophobic
optically transparent self-cleaning surfaces \cite{FraunTransp},
\cite{Transp1}, \cite{Transp2}, which may be used as coatings in the
automotive field, as car-windshields and biker helmets, where impacting rain
drops must be easily repelled. Some studies have shown, indeed, that falling
drops may fully rebound on such water-repellent surfaces with very high
restitution coefficients $\sim0.9$ \cite{QuereRimb}. As a consequence of the
high technological and commercial impact of super-hydrophobic surfaces, in the
last decade a great deal of research has been spent trying to mimic the
super-hydrorepellent surfaces of living organisms. Thus, many artificial
surfaces have been prepared attempting to achieve this objective. Fig.
\ref{Fig_SuperficiArtificiali} shows some examples of super-hydrorepellent
man-made surfaces. At a first sight, different surfaces may appear equally
good candidates to mimic the super-hydrophobic properties of Natural surfaces.
Indeed, being inspired by the solutions offered by Nature, several geometries
have been explored, e.g. uniform arrays of flat topped cylindrical pillars
\cite{callies-cylinders} [Fig. \ref{Fig_SuperficiArtificiali}(a)], or tapered
asperities \cite{Martines-Conical-Shape}, [see Fig.
\ref{Fig_SuperficiArtificiali}(b)], or even nanotube forests with rounded tips
\cite{nanotube-forest} as shown in Fig. \ref{Fig_SuperficiArtificiali}(c).
Beside these few examples, also super-hydrophobic fractal surfaces have been
produced \cite{nanoamphiphil}-\cite{Onda2}, which show apparent contact angles
(CAs) up to $174%
{{}^\circ}%
$. However, the different shapes of such microstructured surface, which
reflect the variety of Natural solutions, should also have some practical
implications which makes them not really completely equivalent.
Liquid drops on super-hydrophobic microstructured surfaces may be
observed mainly in two different states (although intermediate states may also
exists \cite{Moulinet}, \cite{DeSimone}) the Cassie-Baxter \cite{Cassie} and
the Wenzel \cite{Wenzel} states. A drop in a Cassie-Baxter state is just
suspended on the asperities of the underlying surface, which therefore behaves
as a fakir-carpet. The Wenzel state is, instead, characterized by complete
contact between the drop and the substrate. The Wenzel state is usually
unwanted as it results in a strong adherence between the drop and the
substrate and in a very pronounced CA hysteresis \cite{LafunaTransitio},
\cite{LafumaBIS}. Thus, in order to prevent strong adhesion between the drop
and substrate one has to design the super-hydrophobic microstructured surface
in such a way to prevent the Wenzel state to be formed and make stable the
fakir-droplet state. In such sense a first attempt was made in
\cite{NeumannGood} where a model employing the capillary rise of a liquid in
contact with a stripwise heterogeneous surface was developed to study the
effect on contact angles. More recently in \cite{BittounMarmour} two criteria
were proposed to compare the superhydrorepellent properties of different
microstuctures from a wetting point of view. However, the proposed criteria do
not take into account the effect of the internal pressure of the droplet,
which is strictly related to surface tension and curvature of the air-liquid
spherical interface and therefore to its volume. An interesting new approach
with molecular dynamics was proposed in \cite{YangTartaglinoPersson}, to study
the behavior of liquid nanodropltes on rough surfaces. However, as shown by
different authors \cite{LafunaTransitio}, \cite{carbone Idro}, \cite{Bartolo},
\cite{LafumaBIS}, the drop pressure may have a critical role in determining if
composite interface may be formed at the interface between the drop and the
microstructured substrate. Large drop pressures may be generated during the
impact of drops on the substrate, and in this case, as shown in Ref.
\cite{QuereRimb}, \cite{Nosonovsky1}, \cite{Bartolo} high impact velocities
(i.e. large impact pressures) may destabilize the fakir-state, cause the
transition to the Wenzel state and make the droplet not able to bounce.
\begin{figure}[ptb]
\begin{center}
\includegraphics[
height=5.48cm,
width=12.99cm
]{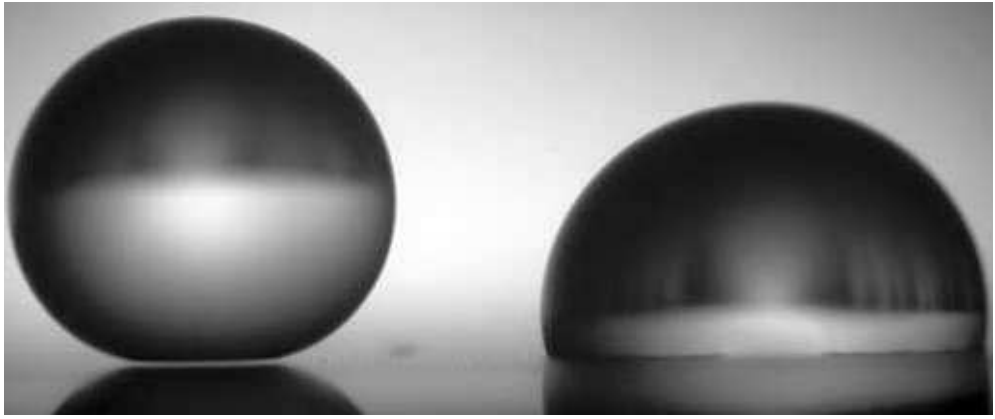}
\end{center}
\caption{Two millimetric water drops of the same volume on a microstructured
super-hydrorepellent surface (adapted from Ref. \cite{callies wenzel-cassie}).
A fakir state (i.e. Cassie-Baxter state) is obtained by gently placing the
drop on the substrate (on the left). A full-contact condition (Wenzel state)
is obtained by incresing the drop pressure, e.g. by let the drop impacing the
surface with enough speed (on the right).}%
\label{FIg. Coesisting Cassie and Wenzel States}%
\end{figure}Fig. \ref{FIg. Coesisting Cassie and Wenzel States} shows indeed
two water drops of the same volume on a microstructured super-hydrorepellent
surface. The drop on left has been gently deposited on the substrate, thus
allowing the formation of a fakir-droplet state, the drop on the right has
instead undergone a transition to the Wenzel state as a consequence of an
increase of the drop pressure above a critical threshold.

The transition between the composite (Cassie) and wetted (Wenzel) states has
been investigated theoretically in many papers (\cite{Nosonovsky2},
\cite{Nosonovsky3}, \cite{Nosonovsky4}, \cite{Nosonovsky5}, \cite{Nosonovsky6}%
, \cite{Patankar1}), but usually the effect of the liquid drop pressure is not
taken into account. Only very recently an interesting study \cite{Patankar2}
has been presented where the superhydrorepellent properties of a surface with
cavities has been investigated and the effective energy of such systems has
been studied by also including the influence of drop pressure. In this paper
we study the system by means of an energy approach of which the most general
formulation has been given by Lipowsky \cite{Lipowsky}, who by means of a
minimization technique proposed a generalization of the Cassie-Baxter and
Wenzel laws for liquid drop sitting on chemically heterogeneous but flat
substrates, where the area fraction of each phase was given a priori. Also,
there are others paper treating the problem of hydraulic pressure in
determining the stability of Cassie-Baxter state. \cite{Zheng} and
\cite{LobatonSalamon}, for example, treated this problem in the case of
pillars with flat tip with sharp edges so that the fraction of the projected
area that is wet is assigned a priori. In particular, \cite{Zheng} gave an
expression of the critical hydraulic pressure for which the transition process
between the Cassie-Baxter wetting mode and the Wenzel one occurs,
\cite{LobatonSalamon} also studied the effect of the contact line length by
varying the shape of the pillars while maintaining constant the area. Indeed
for rounded pillars the wetted area is not fixed a priori and should be
determined as a part of the problem through a minimization technique. In
\cite{LiuLange} the authors studied the case of a surface with an array of
small spheres on it and determined the pressure required to force the meniscus
to bend sufficiently to touch the underlying surface. However, if the pillars
are sufficiently tall the transition to Wenzel state may be achieved before
the meniscus touches the underlying surface because of thermodynamic instability.

In Ref. \cite{carbone Idro} it has been shown that to prevent the impalement
transition it is necessary to increase the critical pressure $p_{W}$ at which
this transition occurs (we refer to $p_{W}$ as the Wenzel pressure); high
values of $p_{W}$ are indeed a strict requirement in those engineering
applications where rain falling drops have to be supported by the substrate.
However, in many cases we also want the liquid drops be very easily detached
from the substrate, i.e. the pull-off pressure $p_{out}$ be reduced almost to
zero. This property is, for example, found in many insects, that usually walk
or skate on the free liquid-air surface. Such insects not only need not to
sink into the water, but also need to detach easily from the liquid free and
surface move and run easily on it.

In a preceding paper \cite{carbone Idro} one of the authors has analyzed the
wetting/non-wetting properties of a liquid drop in contact with an extremely
idealized 1D rough profile, i.e. a simple sinusoidal profile. The analysis
have clarified some theoretical points (mainly from a qualitative point of
view) of wetting non wetting behavior of super-hydrorepellent surfaces. Here
the study is extended to 2D microstructured surfaces with periodic a
distribution of axial-symmetric micropillars, which on the other hand are very
commonly utilized in such applications. Some hints to design such
super-hydrophobic surfaces to achieve both the aim of large $p_{W}$ values and
low pull-off pressures $p_{out}$ are also provided.

\section{Formulation\label{Section Elliptic Equation}}

Let us consider a periodic distribution of chemically hydrophobic (e.g.
fluorinated) pillars (thermodynamic contact angle $\theta_{e}>\pi/2$). We
assume that the elementary cell of the periodic structure is a square,
although we can deal with any type of periodic distribution of micro-pillars.
We develop the analysis for three types of micro-pillars: conical pillars,
hemispherical topped cylindrical pillars, and flat topped cylindrical pillars.
We assume that the micropillars are stiff enough to consider negligible their
deformation under the action of the drop pressure. This, indeed, is a very
good approximation in many practical cases, and is always employed in
theoretical investigations dealing with super-hydrorepellence. We also assume
that the liquid is incompressible (i.e. we neglect the contribution of the
liquid elastic energy) and the drop is very slowly evaporating (i.e. the time
scale to reach the equilibrium is much shorter than the time scale of the
evaporation). We observe that, being the diameter of the liquid drop in the
range of millimeters, whereas the linear spacing $2\lambda$ between the
pillars in the range of micro- or even nano-meters, the drop can be considered
as a semi-infinite liquid space when we analyze the problem at the microscale.
Thus, when we look the drop-substrate interface at very large magnifications
the state of the systems is completely determined by the following state
parameters: the drop-pressure $p$ at the liquid-substrate interface, the real
solid-liquid contact area, the liquid-air free surface at the interface, and
the penetration $\Delta$ (see below) of the liquid drop inside the pillar forest.

\begin{figure}[ptb]
\begin{center}
\includegraphics[
height=6.59cm,
width=12.99cm
]{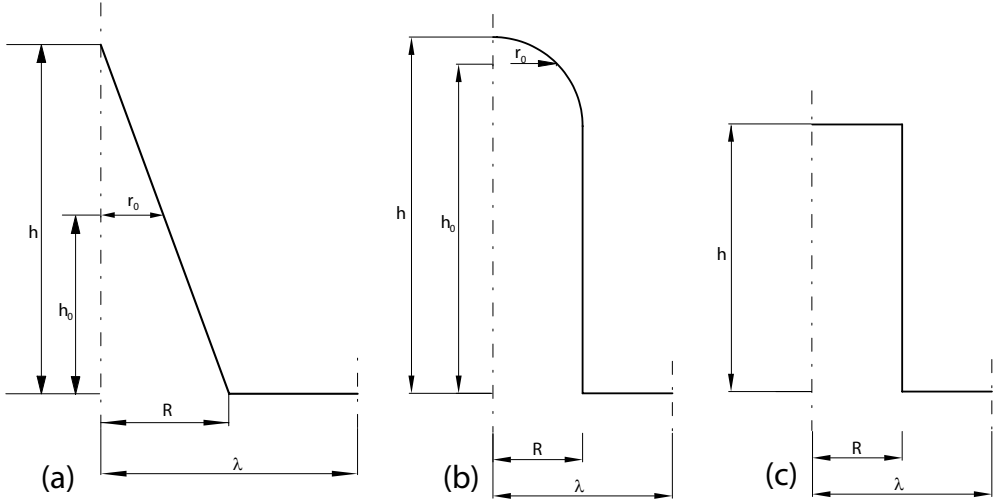}
\end{center}
\caption{Geometry of the conical (a), emispherical topped (b), and cylindrical
(c), pillars.}%
\label{Fig. Geometry Pillars}%
\end{figure}

Fig. \ref{Fig. Geometry Pillars} shows the geometry of the pillars and the
parameters we use to describe the position of the triple line. The reference
plane $(x,y,0)$ is placed at the base of the pillars, and the $z$-coordinate
is directed toward the top of the pillars. The $z$-coordinate of the free
liquid surface will be referred to as $u\left(  x,y\right)  $.
Because of the substrate corrugation, the liquid can either wet
the whole substrate surface or be in stable or metastable partial contact with
it. In case of partial contact the liquid/air interface must satisfy the
Laplace formula. Assuming that the slope of the liquid-air interface is
sufficiently small, which simply requires that $p\lambda/\left(  2\gamma
_{LA}\right)  \ll1$, the Laplace formula reads%
\begin{equation}
\nabla^{2}u\left(  x,y\right)  =u_{xx}\left(  x,y\right)  +u_{yy}\left(
x,y\right)  =\frac{p}{\gamma_{LA}}\label{Laplace2}%
\end{equation}
where $u_{xx}=\partial^{2}u/\partial x^{2}$, $u_{yy}=\partial^{2}u/\partial
y^{2}$ and $\gamma_{LA}$ is the liquid-air surface tension. The condition
$p\lambda/\left(  2\gamma_{LA}\right)  \ll1$ is satisfied in most cases, e.g.
in case of water $\gamma_{LA}=72$~$\mathrm{mJ/m}^{2}$ and assuming
$\lambda\approx1\mathrm{\mu m}$ one obtains $p<1.4~\mathrm{bar}$ which is in
most cases true. This then implies the angle which the free surface forms with
the pillar, for the geometry we have investigated and $R\leq\lambda$ (where
$R$ is the radius of the pillars), need to be less than $30%
{{}^\circ}%
$.

Eq. (\ref{Laplace2}), because of periodicity,\thinspace can be solved over a
quarter of the elementary square cell. However we need also to specify
boundary conditions. A first boundary condition has to be written at the
triple-line, which represents the contour delimiting the liquid-solid
interface. Let us call the projection of this contour on the $(x,y,0)$
reference plane with the symbol $L$. We observe that, in general, the curve
$L$ (for which the mathematical expression in implicit form can be given as
$f_{L}\left(  x,y\right)  =0$) is not known \textit{a priori} and has to be
determined by requiring that the total energy of the system is stationary at
equilibrium (see below). Therefore at the triple line the following equation
must hold true%
\begin{equation}
u\left(  x,y\right)  =h_{0}\left(  x,y\right)  ;\qquad\left(  x,y\right)  \in
L \label{dirichelet cond}%
\end{equation}
where $L=\left\{  \left(  x,y\right)  \in\Re^{2}\mid f_{L}\left(  x,y\right)
=0\right\}  $, and $h_{0}\left(  x,y\right)  $ is the function describing the
shape of the pillars. Eq. (\ref{dirichelet cond}) simply states that the
liquid-air interface and the solid-liquid interface must intersect at the
triple line. Also the following Neumann boundary conditions must be satisfied
to account for periodic conditions%
\begin{align}
u_{x}\left(  0,y>y_{0}\right)   &  =0;\text{ }u_{x}\left(  \lambda,y\right)
=0\nonumber\\
u_{y}\left(  x>x_{0},0\right)   &  =0;\text{ }u_{y}\left(  x,\lambda\right)
=0 \label{Neumann boundary}%
\end{align}
where $u_{x}=\partial u/\partial x$, $u_{y}=\partial u/\partial y$, $x_{0}$
satisfy the condition $f_{L}\left(  x_{0},0\right)  =0$ and, similarly,
$y_{0}$ satisfies the condition $f_{L}\left(  0,y_{0}\right)  =0$.

\begin{figure}[ptb]
\begin{center}
\includegraphics[
height=7.65cm,
width=7.99cm
]{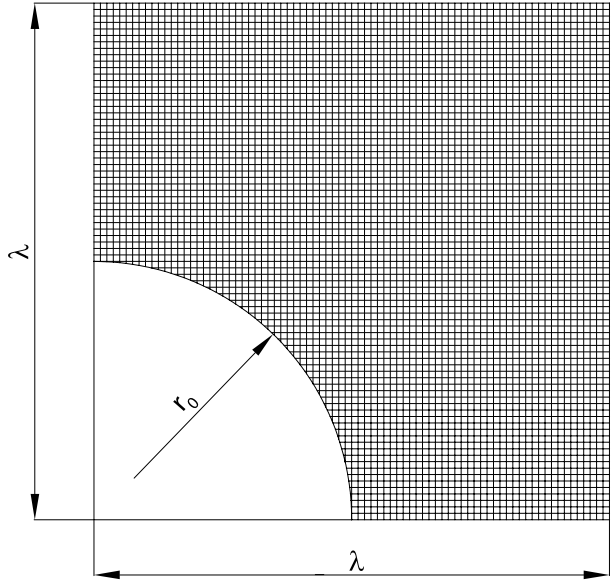}
\end{center}
\caption{The domain of integration of Eq. (\ref{Laplace2}).}%
\label{Fig_Domain of Integration}%
\end{figure}

Observe, that for flat topped cylindrical pillars the function $f_{L}\left(
x,y\right)  $ is known \textit{a priori} being simply $f_{L}\left(
x,y\right)  =x^{2}+y^{2}-R^{2}$, where $R$ is the radius of the pillar. In the
other two cases $f_{L}\left(  x,y\right)  $ is not known \textit{a priori} and
must be determined as a part of the solution of the problem. Indeed, the
physical problem, we are dealing with, belongs to the class of free boundary
problems, and requires an additional condition to achieve the complete
solution. This additional condition is simply the requirement that at
equilibrium, for any given drop pressure $p$, the total energy of the system
(in our case the Gibbs energy $G$) is stationary. Of course, in the general
case, the Gibbs energy is a functional defined on the vector space of
functions $f_{L}\left(  x,y\right)  $, and one should require that it is
stationary at equilibrium to find the Euler-Lagrange equations and determine
the quantity $f_{L}\left(  x,y\right)  $, making the problem belonging to the
class of variational problems. Therefore, the complete solution seems to be
very complicated and expensive from a numerical point of view. However we can
strongly reduce the complexity of the problem if we recall that $p\lambda
/\left(  2\gamma_{LA}\right)  \ll1$, in such a case the slope of the free
liquid-air surface is small. This implies that also the slope of the contour
representing the triple line is small and we conclude that under this
assumption the triple line will only negligibly deviate from a circumference
in case of axial-symmetric pillars (\cite{LobatonSalamon} showed the variation
of the angle that the free surface forms with the pillars, due to the
tortuosity of the triple line, is negligible (being limited to only $3\%$).
Thus, the unknown function $f_{L}\left(  x,y\right)  $ takes a much simpler
form $f_{L}\left(  x,y\right)  =x^{2}+y^{2}-r_{0}^{2}$, where $r_{0}$ is the
unknown radial position of the triple-line, and the total energy of the system
simply becomes a function of the free parameter $r_{0}$. Thus, $r_{0}$ can be
determined by enforcing the equilibrium condition, i.e. requiring that
$\partial G/\partial r_{0}=0$. However, despite the apparent simplicity of Eq.
(\ref{Laplace2}), the particular shape of the domain of integration (see Fig.
\ref{Fig_Domain of Integration}) does not make it possible to obtain a
solution in closed form. We, therefore, have employed a finite difference
approach to solve Eq. (\ref{Laplace2}) with the mixed boundary conditions Eqs.
(\ref{dirichelet cond}) and (\ref{Neumann boundary}). 

For isothermal conditions and constant pressure the Gibbs energy is%
\begin{equation}
G\left(  r_{0},p\right)  =H\left(  r_{0},p\right)  -p\lambda^{2}\Delta\left(
r_{0},p\right)  \label{G}%
\end{equation}
where $\Delta\left(  r_{0},p\right)  $ is the penetration of the rigid
substrate into the semi-infinite liquid and $H$ is the Helmholtz free energy.
The penetration $\Delta$ is given by%
\begin{equation}
\Delta\left(  r_{0},p\right)  =h-s\left(  r_{0},p\right)  \label{D}%
\end{equation}
where $s$ is the average profile of the liquid. Eq. (\ref{D}) represents the
equation of state of the system since it allows to determine one of the three
quantities $\Delta$, $r_{0}$, $p$ once known the other two. To define the
Gibbs energy we first need to express the Helmholtz free energy $H$ which in
our case is just the total surface energy of the system $H\left(
r_{0},p\right)  =\gamma_{LS}S_{LS}+\gamma_{LA}S_{LA}+\gamma_{SA}S_{SA}$, where
$\gamma_{LS}S_{LS}$, $\gamma_{LA}S_{LA}$ and $\gamma_{SA}S_{SA}$ are the
surface energies at the interfaces liquid/solid, liquid/air and solid/air,
respectively. Utilizing the Young's equation $\gamma_{LA}\cos\theta_{e}%
+\gamma_{LS}-\gamma_{SA}=0$, with $\theta_{e}$ being the Young's CA at
equilibrium, the Helmholtz free energy reads%
\begin{equation}
H\left(  r_{0},p\right)  =\gamma_{LA}\left(  S_{LA}-S_{LS}\cos\theta
_{e}\right)  +\gamma_{SA}S_{S} \label{new free energy}%
\end{equation}
and the Gibbs energy becomes%
\begin{equation}
G\left(  r_{0},p\right)  =\gamma_{LA}\left(  S_{LA}-S_{LS}\cos\theta
_{e}\right)  -p\lambda^{2}\Delta+\gamma_{SA}S_{S} \label{G_adim}%
\end{equation}
In Eqs. (\ref{new free energy}) and (\ref{G_adim}) $S_{S}$ represent the total
surface of the substrate over the single square cell. We now require that at
fixed load (i.e. at fixed drop pressure $p$) the Gibbs energy is stationary to
enforce equilibrium conditions and close the system of equations with the
following condition%
\begin{equation}
\frac{\partial G}{\partial r_{0}}=0 \label{equilib condition}%
\end{equation}
Stability or instability of equilibrium can be easily determined by looking at
the sign of $\partial^{2}G/\partial r_{0}^{2}$: Local stability is guaranteed
when the energy has a local minimum, i.e. when $\partial^{2}G/\partial
r_{0}^{2}>0$, whereas instability is detected when $\partial^{2}G/\partial
r_{0}^{2}\leq0$.

\subsection{The apparent contact angle}

At the macroscopic scale the miscrostructure of the substrate is not
observable, therefore at the macro-scale the observer will measure an apparent
liquid solid surface energy $\left(  \gamma_{LS}\right)  _{eff}$ equal to%
\begin{equation}
\left(  \gamma_{LS}\right)  _{eff}=H/\lambda^{2} \label{H2}%
\end{equation}
using Eq. (\ref{G_adim}) $\left(  \gamma_{LS}\right)  _{eff}$ becomes%
\begin{equation}
\left(  \gamma_{LS}\right)  _{eff}=\frac{\gamma_{LA}\left(  S_{LA}-S_{LS}%
\cos\theta\right)  }{\lambda^{2}}+\left(  \gamma_{SA}\right)  _{eff}
\label{glseffective}%
\end{equation}
where we have defined $\left(  \gamma_{SA}\right)  _{eff}=\left(  \gamma
_{SA}S_{S}\right)  /\lambda^{2}$ as the effective solid-air interfacial
energy. At equilibrium the above definition Eq. (\ref{glseffective}) allows to
write a modified Young's equation and evaluate the apparent contact angle
$\theta_{app}$ as%
\begin{equation}
\cos\theta_{app}=-\frac{\left(  \gamma_{LS}\right)  _{eff}-\left(  \gamma
_{SA}\right)  _{eff}}{\gamma_{LA}}=-\frac{S_{LA}-S_{LS}\cos\theta}{\lambda
^{2}} \label{theta1}%
\end{equation}
The above Eq. (\ref{theta1}) represents a generalization of the Cassie-Baxter
equation \cite{Cassie}, which takes into account the influence of the
interfacial drop pressure, and, of course, holds true only at equilibrium.

The above simple considerations make clear that a design criteria based on the
maximization of the apparent contact angle is effective only if we include the
effect of the pressure. Therefore, an optimization of the surface topography
only based on the apparent contact angle evaluated by the Cassie-Baxter
equation can lead to misleading conclusions.

\section{Equilibrium condition: a simplified approach}

The general approach considered above which leads to an optimization procedure
to find Gibbs energy stationary points is very time consuming. For this reason
here we present a much simpler procedure to determine the equilibrium
conditions without the need of a complicated and time-consuming minimization
procedure. Indeed, if we were able to enforce the equilibrium condition before
solving Eq. (\ref{Laplace2}), we would have the advantage of strongly reducing
the complexity of the problem, since, as we are going to show, this allows to
readily calculate the contact radius $r_{0}$ for any given drop pressure $p$
and hence to 'bypass' the minimization of the Gibbs energy. This, indeed, can
be done by requiring that the CA at the triple line is just equal to the
Young's contact angle $\theta_{e}$. Therefore, as shown above, that the triple
line negligibly deviates from a circumference this condition allows to
calculate the real liquid-solid contact area for any given drop pressure. Of
course the determination of the complete thermodynamic state of the system
still requires the calculation of the free liquid-air interface $u\left(
x,y\right)  $ to determine the total interfacial energy and therefore the
apparent contact angle $\theta_{eff}$ and penetration $\Delta$. However, this
calculation can be carried out \textit{a posteriori }by solving Eq.
(\ref{Laplace2}), with the conditions (\ref{dirichelet cond}) and
(\ref{Neumann boundary}), without handling the minimization problem of the
total energy. \begin{figure}[ptb]
\begin{center}
\includegraphics[
height=8.52cm,
width=12.99cm
]{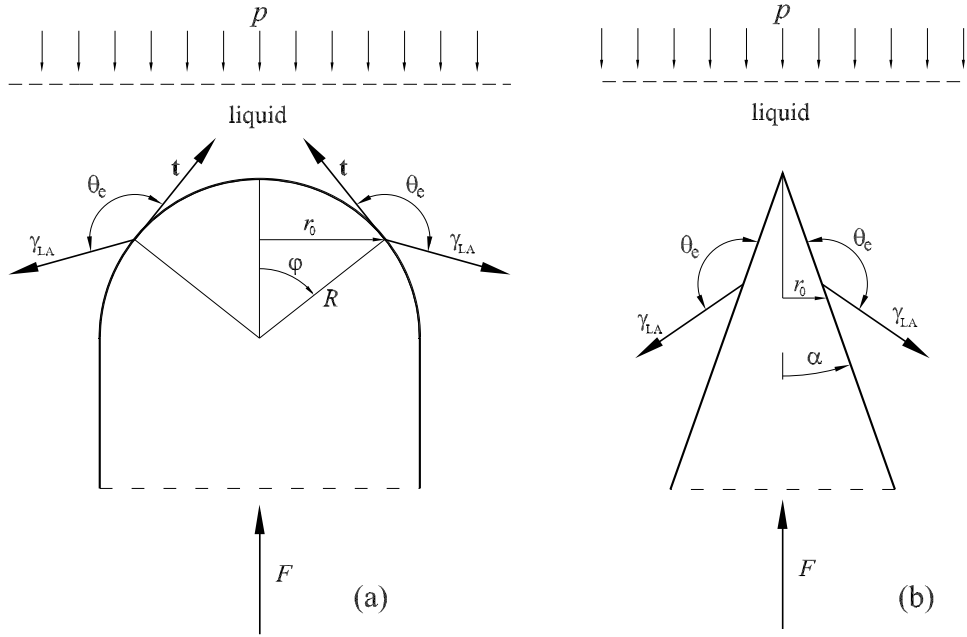}
\end{center}
\caption{The forces acting on the emispherical topped pillars, (a); and on the
conical pillar, (b). Note that we have assumed that the triple line contact is
circumference of radius $r_{0}$ and that the CA angle is just equal to the
Young's contact angle $\theta_{e}$.}%
\label{Fig_Equilibrium_Pillars}%
\end{figure}

\subsection{Conical pillars\label{conical pillar simple apprach}}

Let us observe Fig. \ref{Fig_Equilibrium_Pillars} (b) where the forces acting
on a conical pillar are shown to be the force $F$ that the rigid substrate
applies to the pillar, the surface tension $\gamma_{LA}$ at the triple line,
and the pressure $p$ of the liquid. Thus, the equilibrium writes as%
\begin{equation}
2\pi r_{0}\gamma_{LA}\cos\left(  \theta_{e}-\alpha\right)  +F-\pi r_{0}^{2}p=0
\label{conical equilibrium}%
\end{equation}
where $\alpha$ is the half-cone angle. Now let be $A$ the measure of the area
covered by the elementary cell (not necessarily square) of our periodic
distribution of asperities, (in the case of a square cell the quantity
$\lambda$ is simply $\lambda=A^{1/2}/2$), and enforce the equilibrium of the
whole single cell. Because of the Neumann boundary condition Eq.
(\ref{Neumann boundary}) the liquid-air surface tension at the outer
boundaries of the single cell will not give any contribution to the
equilibrium along the $z$-direction. Therefore we can write $F=pA$ and the
above Eq. (\ref{conical equilibrium}) becomes%
\begin{equation}
\hat{p}=-\frac{\pi\left(  \hat{r}/2\right)  \cos\left(  \theta_{e}%
-\alpha\right)  }{1-\pi\left(  \hat{r}/2\right)  ^{2}}
\label{dimensionless conical eq}%
\end{equation}
where $\hat{r}=r_{0}/\lambda$ and $\hat{p}=p\lambda/\gamma_{LA}$, being
$\lambda=A^{1/2}/2$ a characteristic length. Notice when $\alpha>\theta
_{e}-\pi/2$\ the pressure becomes negative and the drop wets the substrate in
a Wenzel state. Hence the angle $\alpha$ has to satisfy the condition
$0<\alpha<\theta_{e}-\pi/2$ to guarantee a positive value of the drop pressure
$\hat{p}>0$. In such case the drop pressure $\hat{p}$ continuously increases
with the radius $\hat{r}$, i.e. stability of equilibrium is always guaranteed.
The pull-off pressure can be easily evaluated as $\hat{p}_{out}=\hat{p}\left(
\hat{r}=0\right)  $, which as shown by Eq. (\ref{dimensionless conical eq}) is
zero, i.e. drops on conical pillars can be very easily detached from the
substrate. We observe that in all practical cases the conical tip will never
present a real sharp corner. However this does not change our conclusion. The
conical pillar will always guarantee a negligible pull-off pressure, as the
radius of curvature of its rounded tip is always much smaller than radius of
the pillar itself. Moreover, notice that since the effect of the real
extension of the liquid-solid contact area on adhesion between the drop and
surface is already taken into account when we calculate the contribution of
the surface energy to the total energy of the system the detachment of a
liquid drop from a conical pillar requires a vanishing small force even if the
liquid-solid contact area can be larger than in cases of hemispherical or flat
topped pillars.

\subsection{Hemispherical topped cylindrical pillars}

For hemispherical pillars (see Fig. \ref{Fig_Equilibrium_Pillars} (a)), we can
follow the same procedure as outlined above to write the equilibrium%
\begin{equation}
2\pi\left(  R\sin\varphi\right)  \gamma_{LA}\sin\left(  \theta_{e}%
+\varphi\right)  +F-\pi\left(  R\sin\varphi\right)  ^{2}p=0
\label{emispherical equilibrium}%
\end{equation}
where $R$ is the radius of the sphere and $\varphi$ represents the angular
coordinate of the liquid-pillar triple line [see Fig.
\ref{Fig_Equilibrium_Pillars}(a)]. Eq. (\ref{emispherical equilibrium}) can be
conveniently rewritten in a dimensionless form as%
\begin{equation}
\hat{p}=-\frac{\left(  \pi/2\right)  \hat{R}\sin\varphi\sin\left(  \theta
_{e}+\varphi\right)  }{1-\left(  \pi/4\right)  \hat{R}^{2}\sin^{2}\varphi}
\label{dimensionless emispher eq}%
\end{equation}
where we have still used that $\hat{p}=p\lambda/\gamma_{LA}$ and $\hat
{R}=R/\lambda$. At fixed load, stability implies $d\hat{p}/d\varphi>0$ which
is equivalent to have $\partial^{2}G/\partial r_{0}^{2}>0$, whereas unstable
equilibrium conditions are characterized by value of $d\hat{p}/d\varphi\leq0$,
i.e. $\partial^{2}G/\partial r_{0}^{2}\leq0$. Therefore, the threshold value
of pressure at which instability occurs can be determined by enforcing the
condition $d\hat{p}/d\varphi=0$, i.e.%
\begin{equation}
-\pi\hat{R}^{2}\sin\theta_{e}-8\cos\left(  2\varphi\right)  \sin\theta_{e}%
+\pi\hat{R}^{2}\cos\left(  2\varphi\right)  \sin\theta_{e}-8\cos\theta_{e}%
\sin\left(  2\varphi\right)  =0 \label{instability condition}%
\end{equation}

We observe that two unstable conditions can be found in general: the first one
corresponds to the impalement transition to the Wenzel state and is reached
when the pressure increase over a value $\hat{p}_{W}$ (see below), the second
one is achieved when the pressure decreases below a value $\hat{p}_{out}$
which we call pull-off pressure. In particular we stress that the pull-off
pressure $\hat{p}_{out}$ represents the lowest value of the pressure at which
it is still possible to find a stable minimum of the total energy of the
system. If the liquid pressure decreases below this threshold value the liquid
drop detaches from the substrate. However the drop does not detach as a whole,
but rather via the edge propagation of the triple line toward the inner region
of the liquid-pillar. We also observe that from a conceptual point of view,
the pull-off pressure defined in the present paper is analogous to the maximum
detachment force defined in Ref. \cite{Su} for capillary bridges. In fact, in
our case the thermodynamic contact angle is $\theta_{e}>\pi/2$ and in this
case the maximum detaching force for capillary bridges occurs exactly at the
\textit{transition I} point defined in Ref. \cite{Su} (see figure 5 of the
cited paper). Beyond this point stable contact conditions cannot be achieved
and the drop will necessarily detach from the substrate.

As in case of hemispherical topped pillars $0<\varphi<\pi/2$, the equation
$d\hat{p}/d\varphi=0$ gives just one solution $\varphi=\varphi_{out}$, which
corresponds to the pull-off condition $\hat{p}_{out}=\hat{p}\left(
\varphi_{out}\right)  $. The impalement transition to the Wenzel state is
instead obtained at $\varphi=\varphi_{W}=\pi/2$%
\begin{equation}
\hat{p}_{W}=\hat{p}\left(  \varphi_{W}\right)  =-\cos\theta_{e}\frac{\pi
\hat{R}/2}{1-\pi\hat{R}^{2}/4} \label{Wenzel pressure emispehere and falt cyl}%
\end{equation}

\subsection{Flat topped cylindrical pillars}

\begin{figure}[ptb]
\begin{center}
\includegraphics[
height=8.7cm,
width=12.99cm
]{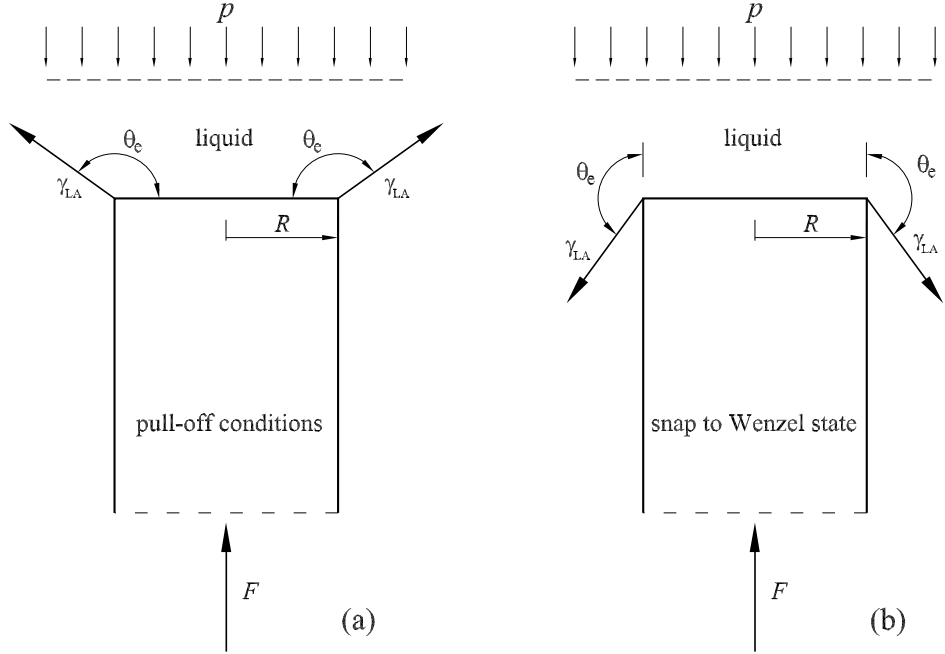}
\end{center}
\caption{The two limiting states that a liquid drop in contact with a
flat-topped cylindrical pillars can assume. Limiting conditions at pull-off,
(a); and limiting condition at the transition from Cassie-Baxter to Wenzel
state, (b).}%
\label{Fig_Cylinders}%
\end{figure}

In case of flat topped cylindrical pillars the liquid-solid contact area on
each pillar is fixed, and equal to $\pi R^{2}$. In such case, depending on the
drop pressure values, the slope of the liquid profile at the triple line can
vary between two different limiting values as clearly shown in Fig.
\ref{Fig_Cylinders}. We can, therefore, easy determine the critical pressure $\hat
{p}_{out}$ at pull-off [see Fig. \ref{Fig_Cylinders}(a)] as%
\begin{equation}
\hat{p}_{out}=-\sin\theta_{e}\frac{\pi\hat{R}/2}{1-\pi\hat{R}^{2}/4}
\label{p out cylinder}%
\end{equation}
and the critical dimensionless pressure $\hat{p}_{W}$ [see Fig.
\ref{Fig_Cylinders}(b)] as (see also \cite{Moulinet})%
\begin{equation}
\hat{p}_{W}=-\cos\theta_{e}\frac{\pi\hat{R}/2}{1-\pi\hat{R}^{2}/4}
\label{Wenzel pressure flat cylinder}%
\end{equation}
which, as expected, is exactly equal to the value found for hemispherical
pillars [see Eq. (\ref{Wenzel pressure emispehere and falt cyl})]. It is
noteworthy that, in case of flat topped cylindrical pillars, the ratio
$\left\vert \hat{p}_{out}/\hat{p}_{W}\right\vert =\tan\theta_{e}$ non
depending on the cylinder radius.

\section{Results}

In what follows we assume that the Young's contact angle is $\theta_{e}=109%
{{}^\circ}%
$, i.e. we assume that the underlying microstructured surface is chemically hydrophobic.

\subsection{Conical pillars\label{sec conical pillar discussion}}

\begin{figure}[ptb]
\begin{center}
\includegraphics[
height=9.82cm,
width=10cm
]{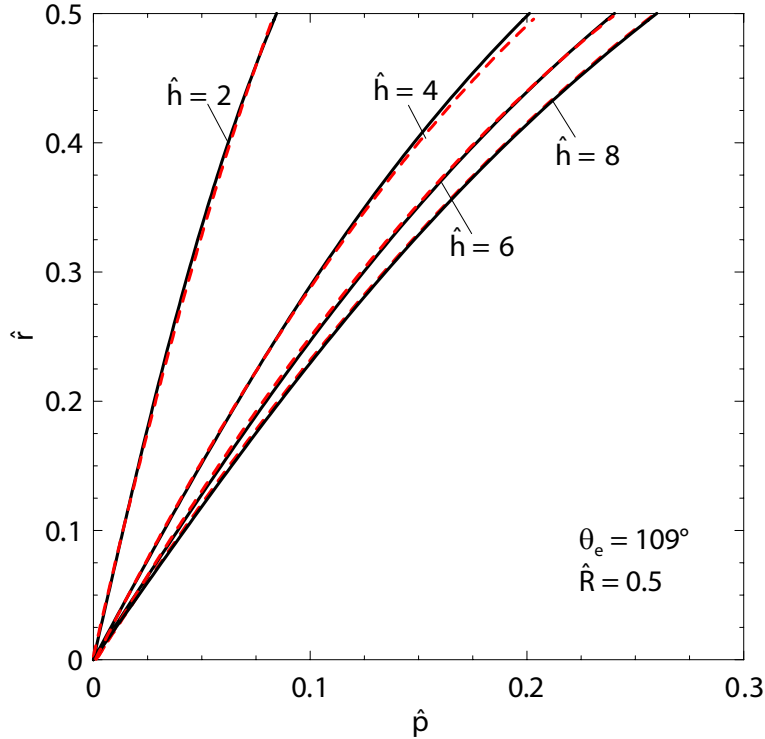}
\end{center}
\caption{The dimensionless radius of contact $\hat{r}$ as a function of the
dimensionless pressure $\hat{p}$ at equilibrium, for a drop in contact with a
periodic distribution of conical pillars. The red-dashed curves are results of
the minimization procedure as explainen in Sec. II, black continuous lines are
instead determined by the simplified approach of Sec. III. The matching
between the two methodologies is excellent. Result are presented for different
values of $\hat{h}=2,~4,~6,~8$, and $\hat{R}=0.5$, the Young's contact angle
is $\theta_{e}=109{{}^{\circ}}$.}%
\label{Comparison old and new procedure}%
\end{figure}

Conical pillars have been analyzed assuming that $\hat{R}=R/\lambda=0.5$,
where $R$ is the radius of the base circle. Fig.
\ref{Comparison old and new procedure} shows the contact radius $\hat{r}$ as a
function of the dimensionless drop pressure $\hat{p}$ for different values of
the dimensionless pillar height $\hat{h}$. The black solid curves are obtained
by Eq. (\ref{dimensionless conical eq}) whereas the red-dashed ones are
obtained by minimizing the total energy of the system as explained in Sec.
\ref{Section Elliptic Equation}. Notice the very good agreement between the
two approaches. Also Fig. \ref{Comparison old and new procedure} shows that
$\hat{r}$ initially increase proportionally to the drop-pressure $\hat{p}$, as
indeed predicted by Eq. (\ref{dimensionless conical eq}), however, as the drop
pressure is increased further, the contact radius also increases and the
denominator in Eq. (\ref{dimensionless conical eq}) may become not negligibly
smaller than one. This, in turn, causes the drop pressure $\hat{p}$ to
increase more than linearly with $\hat{r}$ thus explaining the deviation of
curves in Fig. \ref{Comparison old and new procedure} from linearity.
Since for conical pillars the partial fakir-droplet state is
always stable (i.e. increasing the pressure does not force the system to
undergo a spontaneous transition to the Wenzel state), a different threshold
pressure $\hat{p}_{L}$ has to be defined. This is simply the drop pressure at
which the free liquid-air interface touches for the first time the basal of
the pillars. This condition, indeed, has been shown experimentally to easily
trigger a sharp transition to the Wenzel state \cite{Moulinet}.
\begin{figure}[ptb]
\begin{center}
\includegraphics[
height=9.29cm,
width=10cm
]{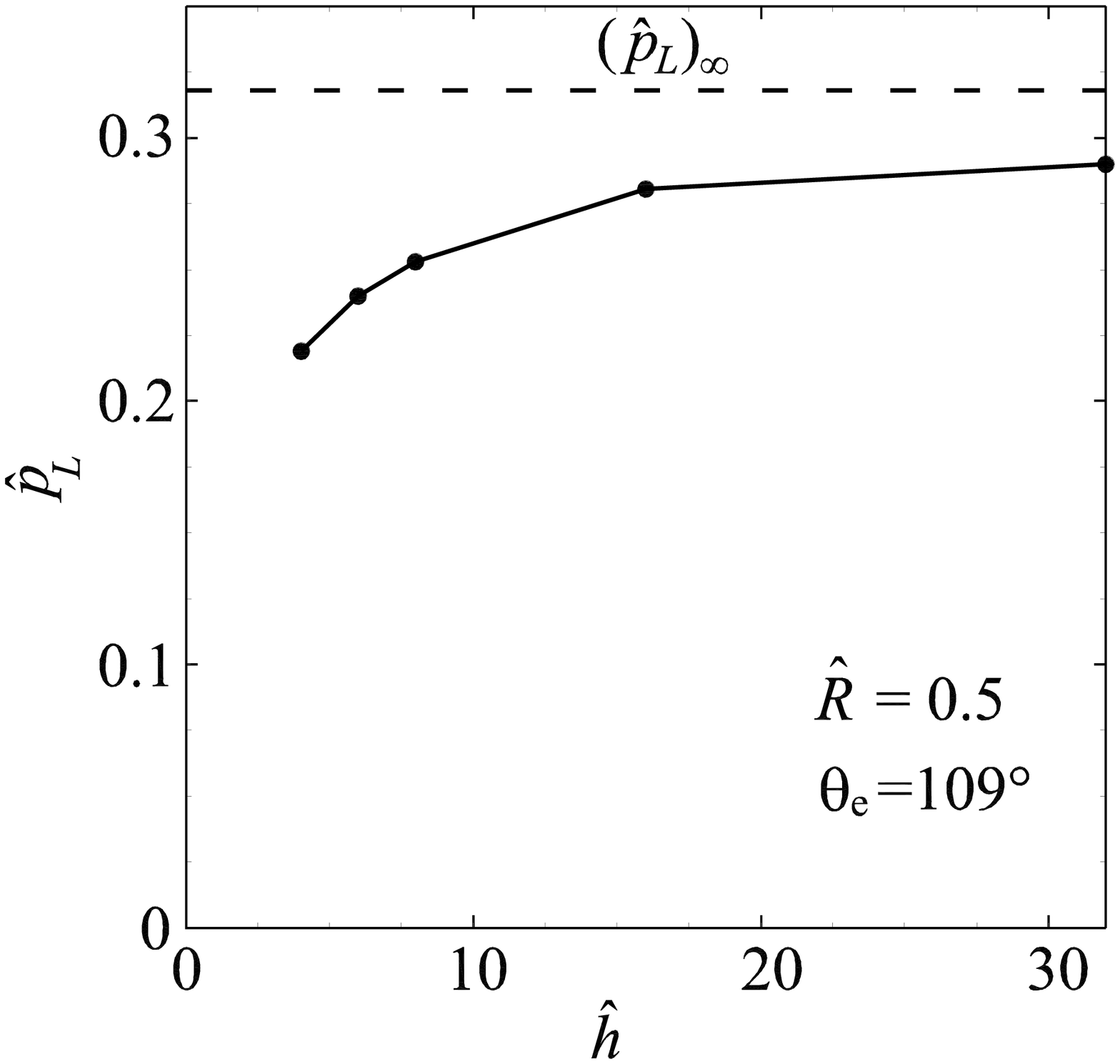}
\end{center}
\caption{The threshold pressure $p_{L}$ as a function of the pillar aspect
ratio $\hat{h}$. Results are presented for $\hat{R}=R/\lambda=0.5$ and
$\theta_{e}=109{{}^{\circ}}.$Observe that increasing $\hat{h}$ the limiting
pressure reaches an asymptotical value $\left(  p_{L}\right)  _{\infty}$ which
is just the values given by Eq. (\ref{Wenzel pressure emispehere and falt cyl}%
).}%
\label{Fig_limiting pressure pl}%
\end{figure}Fig. \ref{Fig_limiting pressure pl} shows the dimensionless
limiting pressure $\hat{p}_{L}$ as a function of the pillar aspect ratio
$\hat{h}$. Observe that increasing $\hat{h}$ also increases the limiting
pressure $\hat{p}_{L}$, since (at fixed $\lambda$) the liquid-air interface
will be farther from the bottom of the pillar. However, $\hat{p}_{L}$ cannot
increase above the limiting asymptotic values obtained for $\hat{h}%
\rightarrow\infty$, $\ $i.e. $\alpha\rightarrow0$. When this happens, the
conical pillar becomes an infinitely tall cylinder of dimensionless radius
$\hat{R}$ for which the Cassie-Baxter state becomes unstable when the
drop-pressure reaches the values $\hat{p}_{W}$ given by Eq.
(\ref{Wenzel pressure flat cylinder}) which indeed is just the asymptotic
value $\left(  \hat{p}_{L}\right)  _{\infty}$ shown by the dashed line in Fig.
\ref{Fig_limiting pressure pl}. \begin{figure}[ptb]
\begin{center}
\includegraphics[
height=10.05cm,
width=10.00cm
]{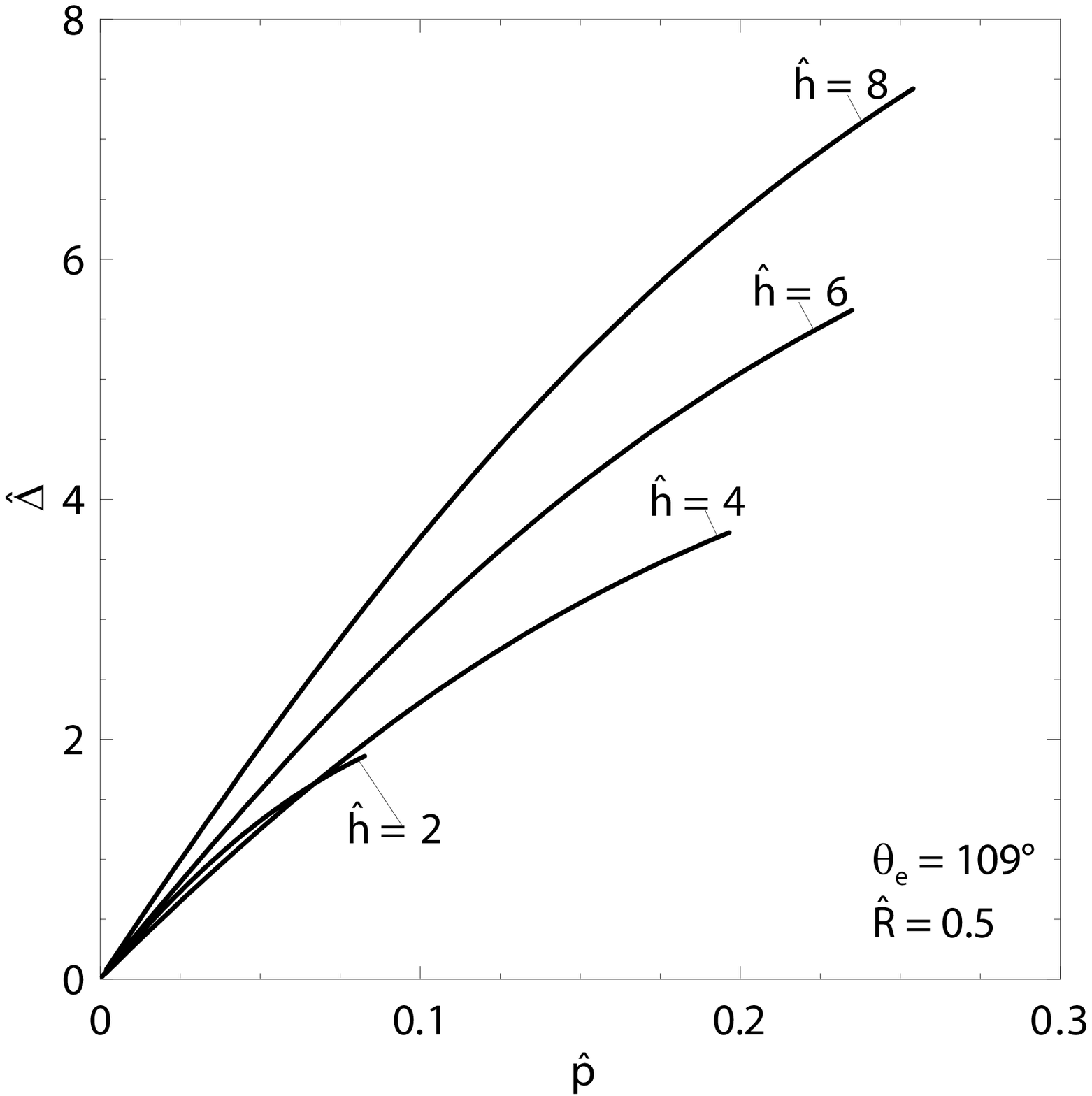}
\end{center}
\caption{The dimensionless drop penetration $\hat{\Delta}$, as a function of
the drop pressure $\hat{p}$. Results are presented for $\hat{R}=0.5$ and
$\theta_{e}=109{{}^{\circ}}$, and for different values of the conical pillars
aspect ratio $\hat{h}$ $=2,~4,~6,~8$.}%
\label{FIgure_Penetration conical pillars}%
\end{figure}

Figure \ref{FIgure_Penetration conical pillars} shows the dimensionless
penetration $\hat{\Delta}=\Delta/\lambda$ of the liquid drop into the pillar
forest as a function of the dimensionless pressure $\hat{p}$. As expected
$\hat{\Delta}$ increases as the pressure $\hat{p}$\ and aspect ratio $\hat{h}$
are increased. Indeed, increasing $h$, at fixed $R$ and $\lambda$, makes the
cone sharper and sharper. This would in turn reduces the liquid-solid contact
radius if the penetration were maintained fixed. But, since the dimensionless
pressure $\hat{p}$ at equilibrium decreases if the contact radius $\hat{r}$ is
reduced [see Eq. (\ref{dimensionless conical eq})], the drop actually needs to
increase the penetration to increase $\hat{r}$ and, thus, sustain the applied
pressure $\hat{p}$.

It is noteworthy to observe that if the dimensionless height of the conical
pillar is strongly reduced, we should expect an increase, instead of a
decrease, of the dimensionless penetration $\hat{\Delta}$ as a function of
$\hat{h}$. Indeed, in such case the half-cone angle $\alpha$ would increase
toward the limiting value $\theta_{e}-\pi/2$ at which the drop spontaneously
undergoes a transition to the Wenzel state (which is obviously characterized
by $\hat{\Delta}=\hat{h}$). This increase of the penetration, as a consequence
of strong reduction of $\hat{h}$, is, indeed, also observed in Fig.
\ref{FIgure_Penetration conical pillars} for $\hat{h}=2$, where the
corresponding $\hat{\Delta}$ values are larger than those obtained for
$\hat{h}=4$ over a large range of $\hat{p}$.

\begin{figure}[ptb]
\begin{center}
\includegraphics[
height=6.61cm,
width=10.00cm
]{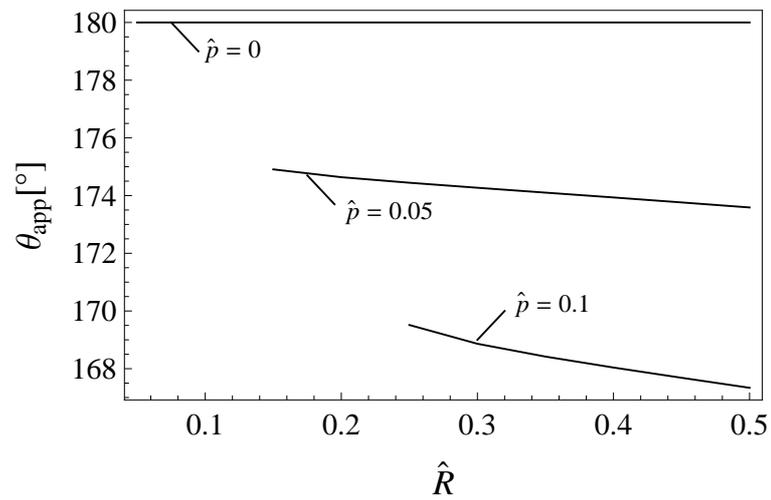}
\end{center}
\caption{The apparent contact angle $\theta_{app}$ as a function of the base
radius $\hat{R}$ for conical pillars. Results are shown for different drop
pressures $\hat{p}$. The aspect ratio of conical pillars is $\hat{h}=8$.}%
\label{Figure_ACA_cone}%
\end{figure}

Figure \ref{Figure_ACA_cone} shows the calculated apparent contact angle
$\theta_{app}$ [see Eq. (\ref{theta1})], as a function of the radius of the
base circle $\hat{R}$ and different drop pressures $\hat{p}$. In the limit
case of zero pressure $\theta_{app}=180%
{{}^\circ}%
$ independently of $\hat{R}$, since in this case the stable state is a perfect
Cassi-Baxter state with the drop just touching the tip of the conical pillars.
Increasing $\hat{p}$ the apparent contact angle decreases in qualitative
agreement with some experimental observations \cite{LafunaTransitio}. The
reason of this decrement is related to the significant increase of penetration
$\Delta$ which in turn determines a strongly increase of the energy term
$p\Delta$ a therefore a reduction of the apparent contact angle [see Eq.
(\ref{theta1})]. Notice that for each given pressure $\hat{p}>0$ a minimum
value of the base radius is needed in order to stabilize the composite
interface and avoid the transition to the Wenzel state. This explains why the
curves at different drop pressures in Fig. \ref{Figure_ACA_cone} are plotted
starting from different values of the base radius $\hat{R}$.

\subsection{Hemispherical topped pillars}

To analyze the behavior of hemispherical topped pillars we use the simplified
approach described in Sec. III to calculate the area of liquid-solid contact
as a function of pressure, then we solve Eq. (\ref{Laplace2}) to determine the
shape of the free liquid-air interface and therefore the penetration of the
liquid drop and the apparent contact angle. \begin{figure}[ptb]
\begin{center}
\includegraphics[
height=7.22cm,
width=15.00cm
]{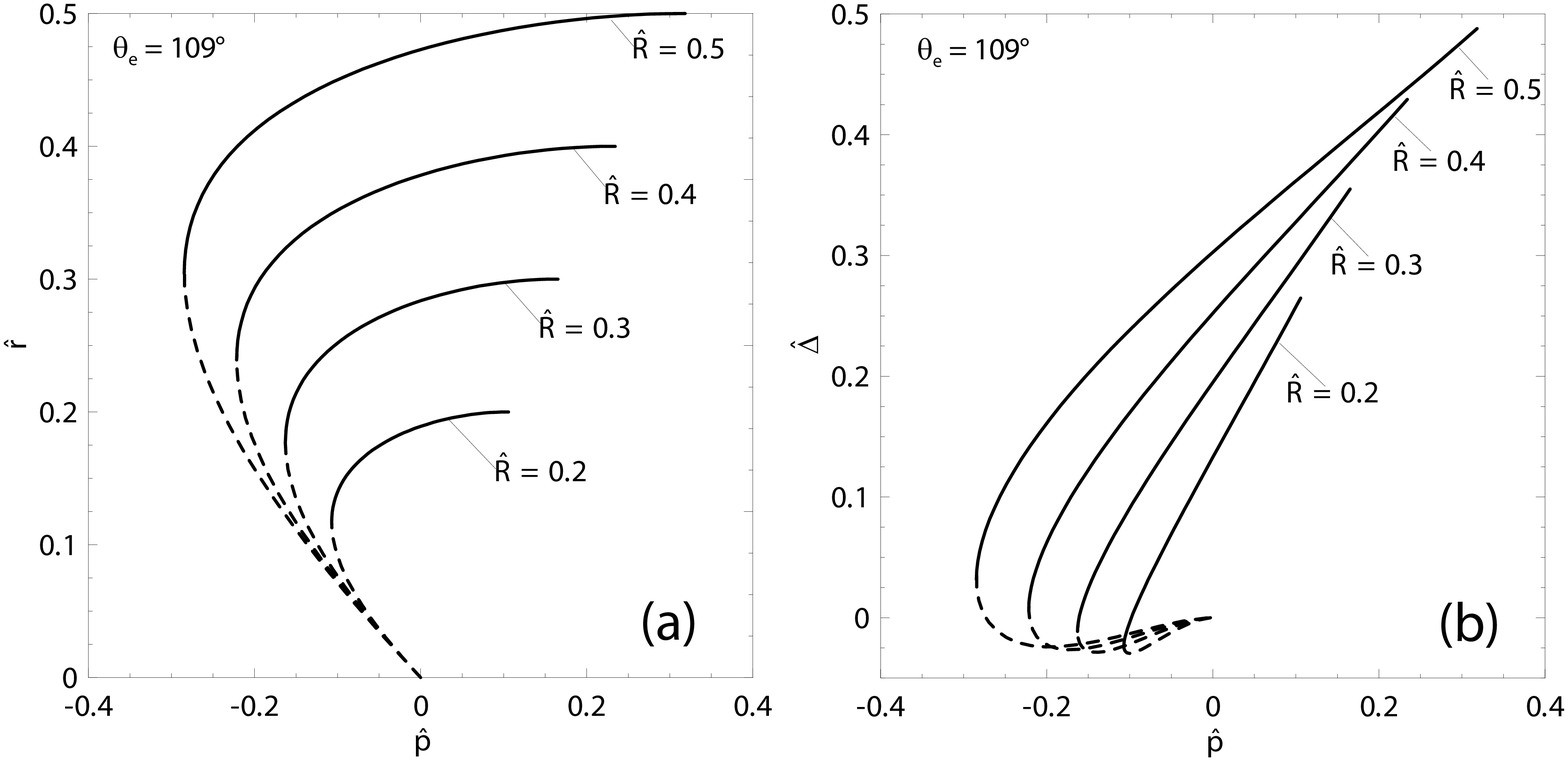}
\end{center}
\caption{The dimensionless radius $\hat{r}$ at equilibrium, (a); and the
dimensionless penetration $\hat{\Delta}$ at equilibrium, (b); as a function of
the dimensionless drop pressure $\hat{p}$ for hemispherical pillars. Full
lines are stable branches, wherease dashed lines are unstable branches at
fixed load (at fixed penetration the stability extends till to the point where
$\hat{\Delta}$ has a minimum). Results are presented for different values of
the dimensionless sphere radius $\hat{R}=0.2,$ $0.3,$~$0.4,$~$0.5$ and for
$\theta_{e}=109{{}^{\circ}}$. Each curve ends at a certain value of $\hat
{p}=\hat{p}_{W}$, which only depends on the pillar radius and contact angle
$\theta_{e}$. When this value of drop pressure is reached the penetration
sharply jumps to the unit value, i.e. the drop ungergoes a sharp transition to
the Wenzel state.}%
\label{Figure Radius and Penetration Emisfere}%
\end{figure}Fig. \ref{Figure Radius and Penetration Emisfere} shows the
dimensionless liquid-pillar contact radius $\hat{r}$ at equilibrium [Fig.
\ref{Figure Radius and Penetration Emisfere} (a)] and the dimensionless
penetration $\hat{\Delta}$ [Fig. \ref{Figure Radius and Penetration Emisfere}
(b)], as a function of the dimensionless drop pressure $\hat{p}$. Two types of
lines are shown. Solid lines are stable equilibrium branches, whereas dashed
lines represent unstable branches at fixed load. As expected, also in this
case the contact radius and the penetration increase with the applied drop
pressure, but this time the $\hat{r}$ vs. $\hat{p}$ law strongly differs from
being linear. Finite negative values of $\left\vert p_{out}\right\vert $ are
related to the finite size of the liquid-solid contact that still exists when
the drop is about to detach from the substrate. This is often strongly
unwanted since, beside the large detaching force, it usually leads also to
large contact angle hysteresis\ \cite{de gennes}, \cite{quere review}. Also
observe that for hemispherical topped cylindrical pillars impalement
transition occurs spontaneously when the drop pressure reaches the limiting
value $p_{W}$ (at which the penetration $\hat{\Delta}$ jumps to $\hat{h}$), in
contrast with what we have found for the conical pillars. Therefore, for
hemispherical pillars the height should be chosen by taking care that at
$p=p_{W}$ the free liquid-air interface does not touch the bottom of the
pillars forest. \begin{figure}[ptb]
\begin{center}
\includegraphics[
height=6.69cm,
width=10.00cm
]{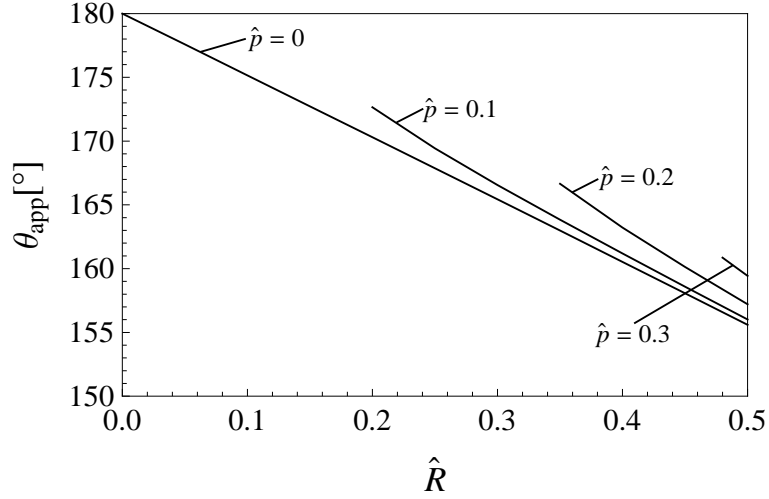}
\end{center}
\caption{The apparent contact angle $\theta_{app}$ as a function of the radius
$\hat{R}$ for hemispherical topped pillars. Results are shown for different
dimensionless drop pressure $\hat{p}$.}%
\label{Figure_ACA_HemiSphere}%
\end{figure}

In case of hemispherical topped pillars, Fig. \ref{Figure_ACA_HemiSphere}
shows that the apparent contact angle slightly increases with pressure
$\hat{p}$ because in this case the liquid/air surface term $\gamma_{LA}S_{LA}$
increases more than the other two energy terms $p\lambda^{2}\Delta$ and
$\gamma_{LS}S_{LS}$. Fig. \ref{Figure_ACA_HemiSphere} also shows the original
Cassie-Baxter solution which corresponds to $\hat{p}=0$.

\subsection{Flat topped cylindrical pillars}

In this case, for any value of the drop-pressure between the two limits
$p_{out}$ given by Eq. (\ref{p out cylinder}) and $p_{W}$ given by Eq.
(\ref{Wenzel pressure flat cylinder}), the contact liquid-pillar area will be
always equal to $\pi R^{2}$ with $R$ the radius of the cylinder. This makes
the problem (\ref{Laplace2}) (\ref{dirichelet cond}) and
(\ref{Neumann boundary}) linear, and, in turn, leads to a direct
proportionality between the drop pressure $p$ and the penetration $\Delta$, as
indeed confirmed experimentally in Ref. \cite{Moulinet}. Notice that, since
$\left\vert p_{out}/p_{W}\right\vert =\left\vert \tan\theta_{e}\right\vert $
and $\theta_{e}$ is usually not larger than $120%
{{}^\circ}%
$, the pull-off pressure is always larger than $p_{W}$. Therefore we expect
that, although a forest of flat topped cylinders can stabilize the
Cassi-Baxter state, a drop suspended on such a microstructured surface is
relatively difficult to detach from it and should suffer of strong contact
angle hysteresis. \begin{figure}[ptb]
\begin{center}
\includegraphics[
height=6.69cm,
width=10.00cm
]{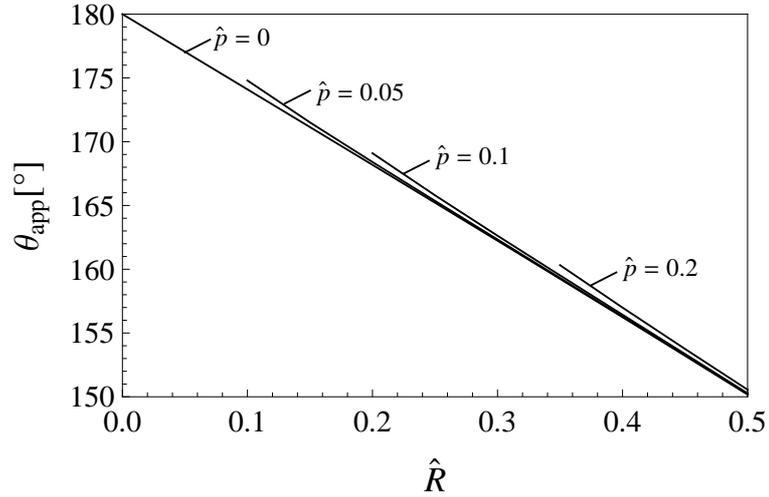}
\end{center}
\caption{The apparent contact angle $\theta_{app}$ as a function of the radius
$\hat{R}$. Results are shown for cylindrical pillars and different
dimensionless drop pressure $\hat{p}$.}%
\label{Figure_ACA_cylinder}%
\end{figure}

Fig. (\ref{Figure_ACA_cylinder}) shows the variation of the apparent contact
angle $\theta_{app}$ as a function of the cylinder radius $\hat{R}$ for
different drop pressures. Similarly to the case of hemispherical pillars the
apparent contact angle grows with the pressure, although in this case this
increment is less pronounced.

\section{Discussion and design suggestions}

Very robust super-hydrorepellent surfaces should possess the ability to
support large drop pressures, and should also allow the drops to easy abandon
the substrate, roll on it with almost zero contact angle hysteresis, or even
easily bounce on it. As already stated, these properties are very desirable in
applications such as micro-fluidic-chips and micro-chemical reactors, where
drops have to be easily moved and positioned. But, also at the macro-scale,
they represent a strict requirement for self-cleaning windows,
water-hydrorepellent windshields, or even hydrorepellent clothing. In these
latter cases, the droplet pressure may reach large values as a consequence of
large inertia forces due to the impact of rain drops. As an example, assuming
the rain drop falls at the speed of about $v_{0}=10$ $\mathrm{m/s}$ we can
easily estimate the maximum impact pressure $p_{\max}\approx\rho v_{0}%
^{2}=1\times10^{5}$ $\mathrm{Pa}$. Therefore, a self-cleaning
super-hydrorepellent window should be necessarily characterized by large
values of the critical Wenzel pressure such that $p_{W}>p_{\max}$. Thus,
assuming that the substrate is constituted of hemispherical topped cylindrical
pillars distribution, and recalling Eq. (\ref{Wenzel pressure flat cylinder})
we found%
\begin{equation}
\lambda<\hat{p}_{W}\frac{\gamma_{LA}}{p_{\max}}=-\cos\theta_{e}\frac{\pi
\hat{R}/2}{1-\pi\hat{R}^{2}/4}\frac{\gamma_{LA}}{p_{\max}}=\lambda_{\max}
\label{impacting drops}%
\end{equation}
So taking for $\hat{R}$ the value $0.5$, recalling that the liquid-water
surface tensions is $\gamma_{LA}\approx72\times10^{-3}$ $\mathrm{J/m}^{2}$,
and assuming a thermodynamic contact angle $\theta_{e}=109%
{{}^\circ}%
$, one ends up with the value $\lambda<230$ $\mathrm{nm}$, i.e. nanotube
forests \cite{nanotube-forest} should be employed in such applications,
provided that the pillar height is sufficiently large to avoid direct contact
between the free liquid-air interface and the bottom of the solid surface (we
also observe incidentally that nanometer spacing between the pillar is also
necessary not to alter the transparency of the glass). However, we underline
that the present paper is focused on steady-state conditions and a correct
treatment for impactig drops should take in account that the curvature of the
free surface is determined by the non uniform pressure distribution at the interface.

In case of conical pillars the critical pressure $p_{L}$ is determined by the
condition that the air-liquid interface touches the pillar forest ground. It
has been observed that the pressure $\hat{p}_{L}$ is always smaller than the
critical Wenzel pressure $\hat{p}_{W}$ calculated for hemispherical and flat
topped cylindrical pillars, and approaches this value only for infinitely
large values of $\hat{h}$. Therefore, one may be tempted to conclude that if
large drop pressures have to be supported, the conical pillar shape is not a
viable solution. However, we can easily correct for this by simply slightly
modifying the pillar design to turn it in a cylinder with a conical tip on the
top, i.e. in a conical topped cylindrical pillar.\ The new conical pillar
design guarantees the same critical Wenzel pressure $\hat{p}_{W}$ as given by
Eqs. (\ref{Wenzel pressure emispehere and falt cyl}) or
(\ref{Wenzel pressure flat cylinder}), but has the fundamental benefit to
present a vanishing or negligibly pull-off pressure $\hat{p}_{out}$. As a
consequence, drop on a forest of conical topped cylindrical pillars should not
suffer from CA hysteresis, as indeed experimentally observed in Ref.
\cite{Martines-Conical-Shape}, and should be able to easily roll or slide on
the substrate or even to bounce on it with very high restitution coefficients.
This, should also explain why some biological systems as water striders, which
usually walk on the free surface of the water, possess a conically shaped
distribution of asperities on their super-hydrorepellent legs \cite{water
striders}. Also note that it is useless to have pillars taller than the
minimum height necessary to prevent the contact of the liquid surface with the
bottom of the pillars forest, in contrast to what is often asserted, i.e. that
the taller the pillars the more the super-hydrorepellence of the surface. The
argument, which is usually provided, is that as the pillar height is
increased, more energy has to be spent to push the drop in full contact with
the substrate. Thus, making pillars taller and taller should lead to a very
large resistance against the impalement transition \cite{quere review}.
However, one should observe that this energy can always be provided by the
pressure $p_{W}$ acting inside the drop independently of the pillar height,
and that the real critical condition for the transition to the Wenzel state to
occur is $p=p_{W}$. Only in case of very small drops (diameter comparable with
the spacing $\lambda$ between the pillars) the physical scenario may slightly
change. In this case, it may be shown that the pillar height can actually play
an additional role: because of mass conservation, an additional resistance
against impalement transition and drop penetration is generated
\cite{Moulinet}.

\section{Conclusions}

In this paper the behaviour of a liquid drop on super-hydrorepellent surfaces
constituted of periodic distribution of pillars has been analyzed. In
particular the critical drop pressure $p_{W}$ which destabilize the
fakir-droplet state causing a transition to a Wenzel (full contact) state and
the critical pressure $p_{out}$ which causes the detachment of the drop from
the substrate have been studied for three types of periodic micro-structured
surfaces: conical, hemispherical topped and flat topped cylindrical pillars,
regularly disposed on a rigid substrate. Both $p_{W}$ and $p_{out}$ are
equally important to assess the superhydrorepellent properties of surfaces. In
fact high $p_{W}$ values are requested in all those applications in which very
high pressures must be supported, e.g. self-cleaning glasses and
super-hydrorepellent windshields, whereas small values of $p_{out}$ are
desirable to guarantee very small contact angle hysteresis and allow the drop
to easily move on the substrate (e.g. microfluidic chemical reactor,
micro-fluidic-chips) or, in case of impacting drops, easily rebound from it
(e.g. self-cleaning windows, super-water-repellent windshields and biker
helmet visors). We have shown that the conical pillars have a pull-off
pressure $p_{out}$ vanishing small (that is an advantage in those applications
in which liquid drops have to be easily removed from the surface), but the
Cassi-Baxter state is destabilized for pressures smaller than the critical
value $p_{W}$ found for hemispherical or flat topped cylindrical pillars.
However, a surface microstructured with cylindrical pillars with conical tips
would have both advantages of large pressure $p_{W}$ and zero pull-off
pressure $p_{out}$. Finally, the effect of the pressure on the apparent
contact angle $\theta_{app}$ has been studied. The analysis has shown that
$\theta_{app}$ reduces significantly with the pressure in case of conical
pillars (in agreement with previous experimental observation), whereas it
slightly increases for hemispherical or flat topped cylindrical pillars.

\begin{acknowledgments}
This work, as part of the European Science Foundation EUROCORES Programme
FANAS was supported from the EC Sixth Framework Programme, under contract N.
ERAS-CT-2003-980409. We kindly acknowledge Dr. Persson for useful comments and discussions.
\end{acknowledgments}


\begin{thebibliography}{99}                                                                                               %


\bibitem {DeSimone}Alberti G., DeSimone A., Wetting of rough surfaces: a
homogenization approach, Proceedings of The Royal Society of London Series
A-Mathematical Physical and Engineering Sciences, \textbf{461} (2053), 79-97 (2005).

\bibitem {Barthlott}Barthlott W., Neinhuis C., Purity of the sacred lotus or
escape from contamination in biological surfaces, Planta \textbf{202}, 1-8 (1997).

\bibitem {BittounMarmour}Bittoun E. and Marmur A., Optimizing
Super-Hydrophobic Surfaces: Criteria for Comparison of Surface Topographies,
Journal of Adhesion Science and Technology, \textbf{23}, 401-411 (2009).

\bibitem {Blossey}Blossey R., Self-cleaning surfaces - virtual realities,
Nature Mater. \textbf{2}, 301 (2003)

\bibitem {callies-cylinders}Callies M., Chen Y., Marty F., Pepin A., Quere D.,
Microfabricated textured surfaces for super-hydrophobicity investigations,
Microelectronics Engineering, \textbf{78-79}, 100-105 (2005).

\bibitem {callies wenzel-cassie}Callies M., Quere D., On water repellency,
Soft Matter, \textbf{1} (1), 55-61 (2005)

\bibitem {Cassie}Cassie A.B.D., and Baxter S., Wettability of porous surfaces,
Trans. Faraday Soc. \textbf{40}, 546-551 (1944)

\bibitem {carbone Idro}Carbone G., Mangialardi L., Hydrophobic properties of a
wavy rough substrate, the European Physical Journal E-Soft Matter \textbf{16}
(1), 67-76 (2005).

\bibitem {FraunTransp}Duparr\`{e} A., Flemming M., Steinert J., and Reihs K.,
Optical coatings with enhanced roughness for ultra-hydrophobic, low scatter
applications, OIC 2001, Banff, Canada, July 15-20 (2001).

\bibitem {nanoamphiphil}Feng L., Li S.H., Li Y.S., Li H.J., Zhang L.J., Zhai
J., Song Y.L., Liu B.Q., Jiang L., Zhu D.B., Super-Hydrophobic Surfaces: From
Natural to Artificial, Adv. Mater., \textbf{14} (24), 1857 (2002)

\bibitem {water striders}Gao X. F., and Jiang L., Water-repellent legs of
water striders, Nature, \textbf{432} (4), 36 (2004).

\bibitem {Gao-Mosquito}Gao X. F., Yan X., Yao X., Xu L., Zhang K., et al, The
dry-style antifogging properties of mosquito compound eyes and artificial
analogues prepared by soft lithography. Adv. Mater. \textbf{19}, 2213--15 (2007)

\bibitem {Hermingauschip}Gau H., Herminghaus S., Lenz P., Lipowsky R., Liquid
Morphologies on Structured Surfaces: From Microchannels to Microchips, Science
\textbf{283}, 46 (1999)

\bibitem {hermin2}Herminghaus S., Roughness-induced non-wetting, Europhys.
Lett., \textbf{52} (2), 165-170 (2000)

\bibitem {de gennes}Joanny J.F., de Gennes P.G., A model for contact angle
hysteresis, Journal of Chemical Physics, \textbf{81} (1), 552-562 (1984)

\bibitem {LafunaTransitio}Lafuma A., and Qu\'{e}r\'{e} D., Superhydrophobic
states, Nature Mater. \textbf{2},\textbf{ }457 (2003)

\bibitem {Lipowsky}Swain P.S. and Lipowsky R., Contact Angles on Heterogeneous
Surfaces: A New Look at Cassie's and Wenzel's Laws, Langmuir \textbf{14},
6772-6780 (1998).

\bibitem {LiuLange}Liu B. and LAnge F.F., Pressure induced transition between
superhydrophobic states: Configuration diagrams and effect of surface feature
size, Journal of Colloid and Interface Science, \textbf{298}, 899-909 (2006).

\bibitem {LobatonSalamon}Lobaton E.J. and Salamon T.R., Computation of
constant mean curvature surfaces: Application to the gas--liquid interface of
a pressurized fluid on a superhydrophobic surface, Journal of Colloid and
Interface Science, \textbf{314}, 184-198 (2007).

\bibitem {nanotube-forest}Lau K. K. S., Bico J., Teo K. B. K., Chhowalla M.,
Amaratunga G. A. J., Milne W. I., McKinley G. H., and Gleason K. K.,
Superhydrophobic Carbon Nanotube Forests, Nano Letters, \textbf{3} (12),
1701-1705 (2003).

\bibitem {Martines-Conical-Shape}Martines E., Seunarine K., Morgan H.,
Gadegaard N., Wilkinson C. D. W., and Riehle M. O., Superhydrophobicity and
Superhydrophilicity of Regular Nanopatterns, Nano Letters, \textbf{5} (10),
2097-2103, (2005)

\bibitem {Moulinet}Moulinet S., and Bartolo D., Life and death of a fakir
droplet: Impalement transitions on superhyrophobic surfaces, Eur. Phys. J. E,
\textbf{24}, 251-260 (2007).

\bibitem {NeumannGood}Neumann A.W. and Good R.J., Thermodynamics of contact
angle, Journal of Colloid and Interface Science, \textbf{38 }(2), 341-358 (1972).

\bibitem {Su}Su Y., Ji B., Huang Y., Hwang K., J. Mater. Sci. \textbf{42},
8885 (2007).

\bibitem {Nosonovsky1}Nosonovsky M. and Bhushan B., Energy transitions in
superhydrophobicity: low adhesion, easy flow and bouncing, J. Phys.: Condens.
Matter \textbf{20}, 395005 (6pp) (2008).

\bibitem {Bartolo}Bartolo D., Bouamrirene F., Verneuil E., Buguin A.,
Silberzan P., Moulinet S., Europhys. Lett., \textbf{74} (2), pp. 299--305 (2006)

\bibitem {Nosonovsky2}Nosonovsky M. and Bhushan B., Biomimetic
Superhydrophobic Surfaces: Multiscale Approach, Nano Letters \textbf{7} (9),
2633-2637 (2007).

\bibitem {Nosonovsky3}Nosonovsky M., Multiscale Roughness and Stability of
Superhydrophobic Biomimetic Interfaces, Langmuir \textbf{23}, 3157-3161 (2007).

\bibitem {Nosonovsky4}Nosonovsky M. and Bhushan B., Roughness-induced
superhydrophobicity: a way to design non-adhesive surfaces, J. Phys.: Condens.
Matter \textbf{20}, 225009 (30pp) (2008).

\bibitem {Nosonovsky5}Bhushan B., Nosonovsky M. and Jung Y. C. Towards
optimization of patterned superhydrophobic surfaces, J. R. Soc. Interface
\textbf{4}, 643-648 (2007).

\bibitem {Nosonovsky6}Nosonovsky M. and Bhushan B., Wetting of rough
three-dimensional superhydrophobic surfaces, Microsyst Technol \textbf{12},
273--281 (2006).

\bibitem {Patankar1}Patankar N., On the Modeling of Hydrophobic Contact Angles
on Rough Surfaces, Langmuir \textbf{19}, 1249-1253 (2003).

\bibitem {Patankar2}Patankar N., Hydrophobicity of Surfaces with Cavities:
Making Hydrophobic Substrates from Hydrophilic Materials?, Journal of Adhesion
Science and Technology \textbf{23}, 413--433 (2009).

\bibitem {Transp1}Nakajima A., Fujishima A., Hashimoto K., and Watanabe T.,
Preparation of Transparent Superhydrophobic Boehmite and Silica Films by
Sublimation of Aluminum Acetylacetonate, Adv. Mater. \textbf{11} (16), 1365 (1999).

\bibitem {Transp2}Nakajima A., Hashimoto K., and Watanabe T., Transparent
Superhydrophobic Thin Films with Self-Cleaning Properties, Langmuir,
\textbf{16}, 7044 (2000).

\bibitem {Onda1}Onda T., Shibuichi S., Satoh N., and Tsujii K.,
Super-Water-Repellent Fractal Surfaces, Langmuir, \textbf{12} (9), 2125-2127 (1996).

\bibitem {QuereIdeas}Qu\'{e}r\'{e} D., Rough ideas on wetting, Physica A
\textbf{313,} 32 (2002)

\bibitem {quere review}Qu\'{e}r\'{e} D., Wetting and Roughness, The Annual
Review of Materials Research, \textbf{38}, 71-99 (2008)

\bibitem {LafumaBIS}Qu\'{e}r\'{e} D., Lafuma A., and Bico J., Slippy and
sticky microtextured solids, Nanotechnology \textbf{14},\textbf{ }1109--1112 (2003)

\bibitem {QuereRimb}Richard D., and Qu\'{e}r\'{e} D., Bouncing water drops,
Europhys. Lett., \textbf{50} (6), 769--775 (2000).

\bibitem {Onda2}Shibuichi S., Onda T., Satoh N., and Tsujii K.,Super
Water-Repellent Surfaces Resulting from Fractal Structure, J. Phys. Chem.,
\textbf{100}, 19512-19517 (1996).

\bibitem {Washizu}Washizu M., Electrostatic Actuation of Liquid Droplets for
Microreactor Applications, IEEE Trans. Ind. Appl., \textbf{34} (4), 732, (1998).

\bibitem {Wenzel}Wenzel R. N., Surface Roughness and Contact Angle, Ind. Eng.
Chem. \textbf{28}, 988-994 (1936)

\bibitem {YangTartaglinoPersson}Yang C., Tartaglino U. and Persson B.J.N.,
Nanodroplets on rough hydrophilic and hydrophobic surfaces, Eur. Phys. J. E,
\textbf{25,} 139-152 (2008).

\bibitem {Zheng}Zheng Q.S., Yu Y. and Zhao Z.H., Effects of Hydraulic Pressure
on the Stability and Transition of Wetting Modes of Superhydrophobic Surfaces,
Langmuir \textbf{21}, 12207-12212 (2005).
\end{thebibliography}
\end{document}